\documentclass[11pt]{article}
\usepackage{geometry}                
\usepackage{graphicx}
\usepackage{amssymb}
\usepackage{epstopdf}
\usepackage{xcolor}
\usepackage{soul}
\usepackage[shortlabels]{enumitem}
\usepackage{multirow}
\usepackage{longtable}
\usepackage{subfigure}

\newcommand{\E}[1]{\mathbb{E}\left[{#1}\right]}
\newcommand{\V}[1]{\mathbb{V}\left[{#1}\right]}

\begin{document}

\title{A Bayesian theory of market impact}

\author{Louis Saddier\\
\small{\'Ecole Normale Sup\'erieure Paris-Saclay, 4 Avenue des Sciences, 91190 Gif-sur-Yvette, France}\\
and \\
Matteo~Marsili\thanks{marsili@ictp.it} \\
    \small{Quantitative Life Sciences Section}\\
    \small{The Abdus Salam International Centre for Theoretical Physics,
  34151 Trieste, Italy}}

\maketitle

\begin{abstract}
The available liquidity at any time in financial markets falls largely short of the typical size of the orders that institutional investors would trade. In order to reduce the impact on prices due to the execution of large orders, traders in financial markets split large orders into a series of smaller ones, which are executed sequentially. The resulting sequence of trades is called a {\em meta-order}. Empirical studies have revealed a non-trivial set of statistical laws on how meta-orders affect prices, which include {\em i)} the square-root behaviour of the expected price variation with the total volume traded, {\em ii)}  its crossover to a linear regime for small volumes, and {\em iii)} a reversion of average prices towards its initial value, after the sequence of trades is over. 
Here we recover this phenomenology within a minimal theoretical framework where the market sets prices by incorporating all information on the direction and speed of trade of the meta-order in a Bayesian manner. The simplicity of this derivation lends further support to the robustness and universality of market impact laws. In particular, it suggests that the square-root impact law originates from the over-estimation of order flows originating from meta-orders. 
\end{abstract}



{D}ata availability makes it possible to probe the laws that govern how financial markets process information to unprecedented precision~\cite{bouchaud2018trades}. Market impact laws are probably the best established empirical observation in high-frequency financial markets~\cite{bouchaud2018trades,prx}. They describe how a {\em meta-order}, which is a sequence of either all buy or sell orders executed by the same trader, affect prices on average. Because of their statistical nature, these laws enjoy a remarkable level of universality, i.e. of independence on details and on context. 
{\color{black}
The aim of this paper is to present a minimal, parameter free theory that accounts for the main features of the observed phenomenology, and delivers falsifiable predictions that suggest ways forward to shed light on the theoretical underpinnings of market microstructure.}

Within market impact phenomenology, the square-root impact law (SRIL) {\color{black} probably stands among the best established empirical observations in high-frequency financial markets}~\cite{bouchaud2018trades,prx}. The SRIL states that a meta-order of total volume $Q$ and duration $T$ changes, on average, the price of the traded asset by an amount~\cite{prx} {\color{black} 
\begin{equation}
\label{empSRIL}
\E{\Delta p_T}\simeq  \pm C\,\sigma_\tau \sqrt{\frac{Q}{V_\tau}}
\end{equation}}
\noindent
where $\Delta p_T=p_T-p_0$ is the price change since the start of the meta-order, $\sigma_\tau$ and $V_\tau$ are the volatility and the volume of transactions measured on the same time-scale $\tau$ (e.g. a day), $C$ is a constant of order one, and the upper (lower) sign holds for a sequence of buy (sell) orders. 

The SRIL enjoys a remarkable universality, because it has been found to hold independently of details, such as the type of asset traded, the market mechanism and the way the sequence of trades is executed (see e.g.~\cite{prx,moro2009market,donier2015million,toth2016square}).
Several proposals to explain the SRIL within mechanistic models have been proposed~\cite{prx,gabaix2003theory,barato2013impact,farmer2013efficiency,donier2015fully,bucci2020co}. Some derive the SRIL from other empirical laws~\cite{gabaix2003theory,farmer2013efficiency} which however appear to have a lower degree of universality. Toth {\em et al.}~\cite{prx} relate the square-root behaviour to the structure of the ``latent order book'', which encodes the propensities of traders to trade at prices close to the market price. This theory~\cite{prx,donier2015fully} predicts that the SRIL holds provided that the density of ``latent'' orders behaves linearly close to the market price. 

Bucci {\em et al.}~\cite{bucci2019crossover} have recently found that the SRIL crosses over to a linear regime for very small $Q$. They relate the crossover to different trading time-scales in the latent order book model. {\color{black} Ref.~\cite{bucci2020co} traces back the origin of the SRIL and of the cross-over to a linear regime to the correlation between different meta-orders. }

For long times, when the sequence of orders is over, the price reverts back towards its original value. 
Because of the scarcity of data for long time-scales, there is no concluding evidence on whether meta-orders leave a permanent impact or not.  The way in which impact decay appears to have a lower degree of universality~\cite{farmer2013efficiency,brokmann2015slow,bucci2018slow,zarinelli2015beyond}. 


This paper aims at deriving this complex phenomenology within a simple theoretical framework based on the Glosten-Milgrom model~\cite{GM} (GMM). The GMM, together with the Kyle model~\cite{kyle1985continuous} and their variants~\cite{back2004information}, is a cornerstone in the literature on market microstructure. This literature investigates how market dynamics emerges from market rules and the strategic behaviour of traders of different types. 

The GMM describes a single-asset market with a population of traders, one of whom is informed about the value of the asset so that she either always sells, if the asset is overpriced, or buys if the asset is underpriced.
In the GMM, the price is fixed by a market maker, based on his expectation of what the value of the asset is, given the past sequence of trades. 
{\color{black} The market maker embodies the collective behaviour of liquidity providers in real markets, who respond to order flows by removing statistical arbitrages.}
As a matter of fact, the informed trader behaves in a deterministic manner, so he trades exactly as a market participant who is executing a {meta-order}. 
In this interpretation, the market maker in the GMM~\cite{GM} provides a stylised description of how markets respond to the statistically persistent order flow generated by meta-orders. 

The key parameter in the GMM is the frequency $\nu$ with which the informed trader submits orders, which translates into the frequency with which child orders of the meta-order are executed. In high-frequency markets there are $\sim 10^4$ transactions a day for a reasonably liquid stock, few of which may be ascribed to a specific meta-order. This implies that market impact laws are related to the limit $\nu\to 0$ in the GMM. 

In this limit, we show that the market impact phenomenology described above is recovered within a fully Bayesian approach where $\nu$ is unknown and is inferred from the sequence of trades. This theory traces the origin of the crossover to a linear behaviour for small $Q$~\cite{bucci2019crossover} to a cutoff in the prior on $\nu$, and it reproduces impact decay. This phenomenology is recovered from a careful correspondence between the GMM and the empirical analysis leading to it. 



We discuss the main steps of the derivation in the following sections and relegate technical details to appendices.

\section{The model}

We consider a financial market populated by many traders, with one asset whose value 
\begin{equation}
\label{EqVt}
W_t=F_t+\theta G
\end{equation}
is composed of two terms: 
$F_t$ is a process with independent increments $F_t-F_{t-1}$ with finite volatility $\V{F_t-F_{t-1}}$.  
The term $\theta G$ is the predictable part of Eq.~(\ref{EqVt}) with $G$ taking values $\pm 1$ with equal probability. $\theta>0$ is a parameter. We call it predictable because one of the traders 
-- the {\em informed} trader -- knows what the value of $G$ is. 
At each time $t=1,2,\ldots$ one trader is drawn at random and she submits an order to buy ($x_t=1$) or sell ($x_t=-1$) one share of the asset. 
With probability $\nu$ the order comes from the informed trader, which means that
$\nu$ is the frequency with which the informed trader submits orders. 
If the asset is underpriced with respect to its value, which occurs if $G=+1$, the informed trader will buy and if it is overpriced -- i.e. if $G=-1$ -- she will sell. 
With probability $1-\nu$ the order is submitted by  uninformed traders, who behave as {\em noise traders}, i.e. they buy ($x_t=1$) with probability $1/2$ and sell ($x_t=-1$) otherwise. 
Hence, the time series $x_{\le t}=(x_1,\ldots,x_t)$ of trades 
is a random vector whose distribution depends on the  value of $G$ and $\nu$, 
\begin{equation}
P\{x_{\le t}|G,\nu\}=\left(\frac{1+G\nu}{2}\right)^{n_t}\left(\frac{1-G\nu}{2}\right)^{t-n_t},~~~~ n_t=\sum_{\tau=1}^t \frac{1+x_\tau}{2}
\label{Pxlet}
\end{equation}
where $n_t$ is the number of buy orders up to time $t$.

The market maker does not know $G$ but he observes the past values $F_{<t}=(F_{0},\ldots,F_{t-1})$ of $F_t$. At time $t$, he fixes the (ask) price $a_t$ at which he will sell and the (bid) price $b_t$ at which he will buy, in a competitive manner\footnote{Traders will be attracted by other market makers if the market maker's expected gain is positive and, if it is negative, he will be driven out of the market.} based on his expectation of the value $W_t$ of the asset, given $F_{<t}$ and the observed sequences of orders $x_{<t}=(x_1,\ldots,x_{t-1})$. He will announce the ask price ${a_{t} = \E{W_t|F_{<t},x_{<t},x_{t}=1}}$ and the bid price ${b_{t} = \E{W_t|F_{<t},x_{<t},x_{t}=-1}}$ for buy orders ($x_t=1$) and sell orders ($x_t=-1$), respectively. The realised price at time $t$ is given by
\begin{equation}
    p_t = \E{W_t|F_{<t},x_{\le t},\nu}=F_{t-1}+\theta\E{G|x_{\le t},\nu}
    \label{ptnu}
\end{equation}
%
where we used the martingale property $\E{F_t|F_{<t}}=F_{t-1}$. 

In the absence of the term $F_t$, the optimal strategy of an informed trader would be to trade as slow as possible, because the expected gain increases as $1/\nu$ as $\nu \to 0$~\cite{touzo2021information}. But in the presence of $F_t$, the price acquires a diffusive increment $\delta F_t=F_t-F_0\sim \sqrt{t}$. The more the informed trader waits between trades, the more the value of her information deteriorates because of the uncertainty about the future value of $F_t$. The optimal range of values of $\nu$ are such that the variation $\delta F_t$ over time intervals $t\sim 1/\nu$ between two trades is of the same order of magnitude of $\theta G$. Therefore we shall assume that 
\begin{equation}
\label{sigmaf}
\sqrt{\V{F_t-F_{t-1}}}=\alpha\theta\sqrt{\nu}\,,
\end{equation}
with $\alpha>0$ a constant of order one\footnote{Eq.~(\ref{sigmaf}) suggests that the stronger the signal $\theta$ that the informed trader receives, the slower she should trade. This seems paradoxical but it is fully consistent with the fact that the profit of informed traders increase the slower they trade, as remarked above. Furthermore, an inverse relation between $\nu$ and $\theta$ is consistent with the fact that $\theta$ quantifies the strength of the market response to persistent order flows (see Eq.~\ref{ptnu}): an informed trader should trade slower the stronger the response of the market.}. We remark that this setting assumes that the market maker knows both $\nu$ and $\theta$.

\subsection{The GMM as a model of market impact}

The informed trader will observe a mismatch $\E{W_t}-p_t=\theta (G-\E{G|x_{\le t},\nu})$ between her expectation of the value of the asset and the market value. As stated before, she will buy if the asset is underpriced ($\E{W_t}>p_t$) which occurs if $G=+1$, and she will sell if it is overpriced ($G=-1$).

To all practical purposes, an informed trader in the GMM is identical to a trader who mechanically executes a buy ($G=+1$) or sell ($G=-1$) meta-order, which is a sequence of buy (or sell) child orders\footnote{This implicitly assumes that the meta-order executer does not behave strategically.}. 
The market maker should set prices as in Eq.~(\ref{ptnu}) irrespective of whether the order flow originates from informed traders or from the execution of a meta-order. A meta-order is defined by a direction $G=\pm 1$ (buy or sell), a volume $Q$ and a time horizon $T$ to execute it which is equal to the traded volume $V_T=T$ over time $T$. The parameter $\nu=Q/V_T$ is the frequency with which child orders are submitted, and is called participation rate in the market microstructure literature (see e.g~\cite{bucci2018slow,zarinelli2015beyond}). For the same meta-order size, $\nu$ is inversely proportional to $V_T$, hence $\nu$ is also a (inverse) measure of liquidity. 
The parameter $\theta$, which is the offset of the value of the asset perceived by the informed trader in the original GMM, tunes the strength of market response to the persistent order flow (see Eq.~\ref{ptnu}) when the GMM is interpreted as a model of market impact.

\subsection{The continuous time limit}

Motivated by the fact that trading in modern financial markets  occurs at the millisecond time-scale, we investigate the scaling limit $\nu\to 0$, $t\to\infty$ with  $q=\nu t$ finite. The rescaled time measures time in terms of meta-order execution and it corresponds to empirical averages conditioned on meta-order size, as in Eq.~(\ref{empSRIL}) for $q=Q$.
The limit $\nu\to 0$ corresponds both to extremely slow speed of meta-order execution and to the limit of perfectly liquid stocks, whereby an infinite number of transactions takes place between two child-orders. We shall express our result in continuous time $q$ while the derivations will be carried out in terms of the microscopic time $t$. 

The volume $V_\tau$ in Eq.~(\ref{empSRIL}) over finite time intervals $q$ is inversely proportional to $\nu$ whereas the volatility, using Eq.~(\ref{sigmaf}), is given by 
\begin{equation}
\label{sigmatau}
\sigma_\tau=\sqrt{\V{F_{t+\tau}-F_t}}=\alpha\theta\sqrt{q}\,,\qquad q=\nu\tau
\end{equation}
The reason why the term $\theta\E{G|x_{\le t},\nu}$ in Eq.~(\ref{ptnu}) does not contribute to the volatility is that $\sigma_\tau$  is computed by an unconditional average wherein the effect of meta-orders is averaged out. By contrast, the expected value in Eq.~(\ref{empSRIL}) is conditioned on the start of the meta-order.

Eq.~(\ref{sigmatau}) for $q=1$ implies that $\alpha\theta=\sigma_{\tau=1/\nu}$ is the volatility over time differences between two child-orders. Therefore $\theta$ depends on the meta-order execution protocol\footnote{Notice that $\sigma_\tau/\sqrt{V_\tau}=\alpha\theta\sqrt{\nu}$ is an empirically observable quantities independent of meta-order execution, and $\alpha$ is a constant. So the product $\theta\sqrt{\nu}$ is independent of the execution protocol.}. 
The universality of market impact laws, such as Eq.~(\ref{empSRIL}), within the idealised setting of the model, implies that results should be expressed in terms of empirically observable quantities, and that they should be independent of $\theta$ (and $\nu$). 



Finally, in the high frequency regime ($\nu\to 0$) it makes sense to specialise to the case where noise traders buy or sell with the same probability, as postulated, and to assume that meta-orders are equally likely to be in either direction, i.e. $P\{G=\pm 1\}=1/2$. 

\subsection{The market impact in the GMM}

The market impact, which is the key quantity of interest to us, is the absolute expected price change since the start of the meta-order\footnote{The literature (see e.g.~\cite{zarinelli2015beyond}) distinguishes the {\em immediate} market impact $\Im(q)$ (for $q<Q$) from the {\em temporary} one (when $q>Q$), and the {\em peak} market impact $\Im(Q)$ when $q=Q$.}
\begin{equation}
\label{Imt}
\Im(q)=\left|\E{\Delta p_t}\right|\,,\qquad q=\nu t
\end{equation}
where $\Delta p_t=p_t-p_0$ and $p_0=F_0$ is the price at the start of the meta-order. It is important to stress that the market impact does not depend on $F_t$.
The expected value in Eq.~(\ref{Imt}) 
corresponds to the statistical average that is performed in the empirical analysis of market impact. This average is conditional to the inception of a meta-order at $t=0$. 
This procedure averages over many different market conditions, thus weakening the long range auto-correlation in the order flow~\cite{bouchaud2018trades}. This justifies, in principle, the uncorrelated order flow generated by noise traders assumed in the model. Accordingly, our analysis focuses on single meta-order statistics and glosses over the effects of correlations between different meta-orders that have been suggested to contribute to market impact~\cite{bucci2020co}.

Note that, by Eq.~(\ref{Pxlet}), the imbalance between buy and sell orders grows linearly in time, i.e. $\E{2n_t-t}= \nu t G$, as long as the meta-order is active ($t\le T$). For small $\nu$ we find (see Appendix~\ref{app:lin})
\begin{equation}
\label{linpt}
p_t\simeq F_{t-1}+\theta G\tanh(\nu^2 t+\nu\sqrt{t}\xi) \,,\qquad (t\le T)
\end{equation}
where $\xi$ is a Gaussian random variable with zero mean and variance one.
This shows that the true value of $G$ would be revealed over times of order $t\sim\nu^{-2}$, which are much longer than the time of the execution of a meta-order if $Q\ll \nu^{-1}$. For small $\nu$ and $t\ll \nu^{-2}$, the market impact $\Im(q)\simeq\theta\nu q$ grows linear in time for $q\le Q$, and it can be cast in the form
\begin{equation}
\label{linImpact}
\Im(q)\simeq \frac{1}{\alpha^2 \theta}\sigma^2_\tau \frac{q}{V_\tau}
\end{equation}
where we used Eq.~(\ref{sigmatau}).
For $q\ll \nu^{-1}$ the impact $\Im(q)$ is negligible with respect to the fluctuations of $F_t$, which are of order $\sqrt{q}$. Hence, the scaling regime of interest $q\ll \nu^{-1}$ lies in a region where the contribution from the meta-order is statistically undetectable.

After the meta-order stops, the statistics of $n_t$ changes because the imbalance of buy and sell orders vanishes and $\E{2n_t-t}= GQ$ remains constant. 
This leads to an apparent permanent impact for times $q\ll \nu^{-1}$ that crosses over to a slow impact decay  ${\Im(q)\simeq \sqrt{\frac{2}{\pi}}\frac{\sigma_\tau}{\alpha\sqrt{V_\tau}}\frac{Q}{\sqrt{q}}}$ for times $q\gg\nu^{-1}$ (see Appendix~\ref{app:lin}). 
Both a linear impact for $q\le Q$ and a slow impact decay for $q>Q$ contradict empirical evidence. Furthermore, the market impact is not independent of meta-order execution, as in Eq.~(\ref{empSRIL}), because of the dependence on $\theta$ in Eq.~(\ref{linImpact}).

\section{A Bayesian market maker}

So far we assumed that the trading frequency $\nu$ of the meta-order is known to the market maker. If the market maker does not know the true value of $\nu$, we can assume that he will update his belief on $\nu$ using Bayes rule and the stream of transactions $x_{\le t}$ he observed 
\begin{eqnarray}
p(v|x_{\le t}) & = & \frac{P(x_{\le t}|v)\phi(v)}{\int_0^1 P(x_{\le t}|y)\phi(y)dy} 
\end{eqnarray}
where $\phi(v)$ is the prior. Here and henceforth, $v$ denotes the inferred value of the  trading frequency of the meta-order (which is a random variable), whereas $\nu$ denotes the true value.
Using this, the market maker estimates the last term in Eq.~(\ref{ptnu}) as
\begin{equation}
\label{ptnt}
\E{G|x_{\le t}}=\int_0^1\!dv\, \E{G|x_{\le t},v}P(v|x_{\le t})
\end{equation}
where $\E{G|x_{\le t},v}$ is the same as in Eq.~(\ref{ptnu}) with $\nu$ replaced by $v$. We remark that the expectation in Eq.~(\ref{ptnt}) as well as the probability $P(v|x_{\le t})$ refer to the subjective state of knowledge of the market maker, not to objective probability distributions. 

\subsection{The square-root-impact law}

\begin{figure}[!ht]
\centering
\includegraphics[width=0.8\textwidth]
{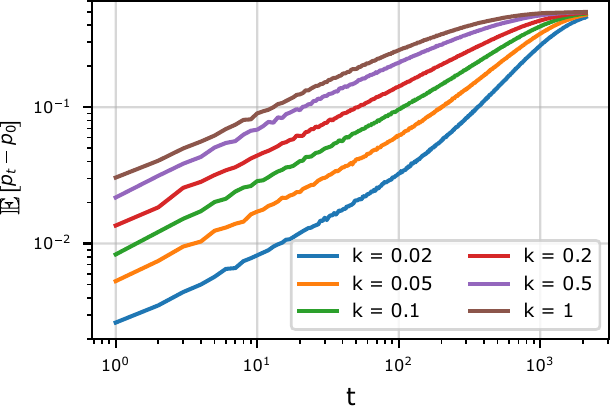}
\caption{Numerical simulations ($\nu=0.1$ and $Y=1$) of price impact for priors of type $\phi(v)\sim A v^{k-1}$. The case $k=1$ is the same as described in Eq.~(\ref{Epterf}). One can remark that other curves have the same slope as this one for $t\sim1/\nu$, emphasising the fact that they also follow the SRIL.}
\label{fig:SRILprior}
\end{figure}

Eq.~(\ref{ptnt}) is more conveniently expressed in terms of the random variable $\xi=(2n_t-t)/\sqrt{t}$, which  asymptotically follows the normal law in the scaling regime $1\ll t\ll \nu^{-2}$, with mean $\E{\xi}=G\,\nu\sqrt{t}$ and variance $\V{\xi}=1-\nu^2$. As shown in  Appendix~\ref{app:sril}, to leading order, 
\begin{equation}
p_t(\xi) \simeq  F_{t-1}+\frac{\theta}{\sqrt{2\pi}}\int_0^\infty\!d\gamma  e^{-\frac{1}{2}(\xi-\gamma)^2}=F_{t-1}+\theta\, {\rm erf}\!\left(\xi/\sqrt{2}\right)\,.
 \label{ptxi}
\end{equation}
Notice that, in the limit $t\to\infty$, the price converges to the true value asymptotically, i.e. $p_t\to F_{t-1}+\theta G$, because $G\xi\to\infty$ in the limit.
Taking the expectation over $\xi$, we find
\begin{equation}
\label{Epterf}
\E{\Delta p_t}\simeq \theta G\,{\rm erf}\!\left(\nu\sqrt{t}/{2}\right) \simeq \theta\frac{G}{\sqrt{\pi}}\nu\sqrt{t}\,.
\end{equation}
In continuous time, using Eq.~(\ref{sigmatau}) the market impact can be expressed as
\begin{equation}
\label{theoSRIL}
\Im(q)=\frac{1}{\sqrt{\pi}\alpha} \sigma_{\tau}\sqrt{\frac{q}{V_\tau}}
\end{equation}
which is Eq.~(\ref{empSRIL}) with $C=1/(\sqrt{\pi}\alpha)$ when $q=Q$. 

The true value of $\nu$ becomes statistically detectable when $q\sim \nu^{-1}$, which is also the time when the true value of $G$ is revealed (see Eq.~\ref{Epterf}). 

It is worth to mention that the volatility $\sigma^2_\tau$ differs from the variance 
\begin{equation}
\label{variance_mart}
\V{\Delta p_{\tau}}=\sigma_\tau^2+\frac{\theta^2}{3}
\end{equation}
of the price difference $\Delta p_\tau=p_\tau-p_0$, because conditioning on the start of the meta-order introduces a constant term due to market impact (see Appendix~\ref{app:sril}). 
This is a prediction that can be tested in empirical analysis.

The derivation of the square-root law outlined above holds under considerably more general conditions than those assumed so far. 
Fig.~\ref{fig:SRILprior} shows that the leading behaviour in Eq.~(\ref{Epterf})  holds generally for any prior with $\phi(v)\sim v^{k-1}$ with $k>0$, as shown in Appendix~\ref{app:prior}\footnote{Zarinelli {\em et al}~\cite{zarinelli2015beyond} estimate that the distribution of the participation rate $\nu$ has a power law distribution with exponent $\kappa\simeq 0.864$.}. The square-root law holds as long as the volume of individual orders is drawn from a distribution with finite second moment and/or in the presence of short range auto-correlations in the orders of noise traders (see Appendix~\ref{app:vol}). 

\subsection{The crossover to short time linear impact}

\begin{figure}[!ht]
\centering
\includegraphics[width=0.8\textwidth]
{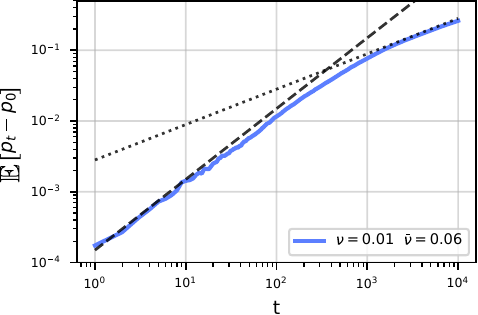}
\caption{Market impact in the GMM by choosing a cutoff $\bar\nu$ in the prior for a buy metaorder ($Y=1$). Blue curve is derived from Monte-Carlo numerical simulations. Dashed line represents the theoretical linear impact we get from Eq.~(\ref{linear}) and dotted line represents the SRIL.}
\label{fig2}
\end{figure}

Eq.~(\ref{Epterf}) assumes a prior $\phi(\nu)$ that extends to finite values of $\nu$. It makes sense to assume that the market anticipates that the contribution of meta-orders to the order flow is small, i.e. that $\nu$ is at most $\bar\nu$. We take $\phi(v)=1/\bar \nu$ for $v\in[0,\bar \nu]$ and $\phi(v)=0$ otherwise in order to incorporate this observation. With this assumption, we find (see Appendix~\ref{app:cross})
\begin{equation}
\label{ptnubar}
\E{G|x_{\le t}} \simeq 2\frac{{\rm erf}\!\left(\frac{\xi}{\sqrt{2}}\right)+{\rm erf}\!\left(\frac{\bar\nu\sqrt{t}-\xi}{\sqrt{2}}\right)}{{\rm erf}\!\left(\frac{\bar\nu\sqrt{t}+\xi}{\sqrt{2}}\right)+{\rm erf}\!\left(\frac{\bar\nu\sqrt{t}-\xi}{\sqrt{2}}\right)}-1
\end{equation}
where $\xi=(2n_t-t)/\sqrt{t}$ has the same definition as above.
For $\bar\nu\sqrt{t}\gg 1$ (i.e. $t\gg\bar\nu^{-2}$) we recover Eq.~(\ref{ptxi}) whereas for $t\ll\bar\nu^{-2}$ the expansion of the expression above for $\bar\nu\sqrt{t}\ll 1$ leads to $\E{G|x_{\le t}}\simeq \frac 1 2 \xi\bar\nu\sqrt{t}+O(\bar\nu^2 t)$. Taking the expectation on $\xi$ we recover the linear regime found in Ref.~\cite{bucci2019crossover}
\begin{equation}
\label{linear}
\Im(q)\simeq \frac 1 2 \theta\bar\nu q+\ldots
~~~(q\ll\nu/\bar\nu^{2})
\end{equation}
In other words, for small orders ($q\ll \nu/\bar\nu^{2}$) we find a linear impact. For orders that last a time $\nu/\bar\nu^{2}\ll q\ll \nu^{-1}$ the SRIL sets in and for orders that are much longer than $\nu^{-1}$ the impact saturates to a finite value. Notice that the crossover value $q^*=\frac{4}{\pi}\nu/\bar\nu^2$ obtained equating 
Eqs.~(\ref{theoSRIL}) and (\ref{linear}) depends on meta-order execution.

For $q=Q$ Eq.~(\ref{linear}) reproduces a crossover of $\Im(Q)$ to a linear behaviour similar to that observed by Bucci {\em et al.}~\cite{bucci2019crossover}, but not identical: while our theory predicts a crossover in terms of meta-order duration (with a crossover time $T^*=\bar\nu^{-2}$), the results in Bucci {\em et al.} suggest that the crossover variable is $\nu\sqrt{T}$ (see Fig. 2 in~\cite{bucci2019crossover}). The picture that emerges from empirical studies is certainly more complex than the simplified picture of the crossover provided by our model. For example, our approach neglects the effect of correlations between meta-orders, which has been suggested to be relevant for the crossover~\cite{bucci2020co}.


\subsection{Impact decay}

Let us now consider the behaviour of market impact for times $t>T$ after the meta-order has been fully executed. Again we can use Eq.~(\ref{ptxi}), but the statistics of $\xi=(2n_t-t)/\sqrt{t}$ should take into account that $\E{x_\tau}=0$ for $\tau>T$. Therefore 
\begin{equation}
\label{eq:expec_buys_decay}
\E{n_t}=\frac t 2 +\frac 1 2 \sum_{\tau=1}^{Q/\nu}\E{x_\tau}+\frac 1 2 \sum_{\tau=Q/\nu+1}^t\E{x_\tau}=\frac t 2 +G \frac Q 2
\end{equation}
The variance $\V{n_t}=\frac t 4 -\frac{\nu Q}{4}$ is computed in a similar manner. Hence $\xi$ is a Gaussian variable with mean $\E{\xi}=G \frac{Q}{\sqrt{t}}$ and variance $\V{\xi}=1-\nu Q/t\simeq 1$. Using this in Eq.~(\ref{ptxi}) one gets $\E{\Delta p_t}
\simeq \theta G\, {\rm erf}\!\left(\frac{Q}{2\sqrt{t}}\right)$ that implies a decay of the impact as
\begin{equation}
\label{eq:exp_price_decay}
\Im(q)\simeq \frac{1}{\sqrt{\pi}\alpha}\sigma_\tau \frac{Q}{\sqrt{V_\tau q}}\,.
\end{equation}

\begin{figure}[!ht]
\centering
\includegraphics[width=0.8\textwidth]
{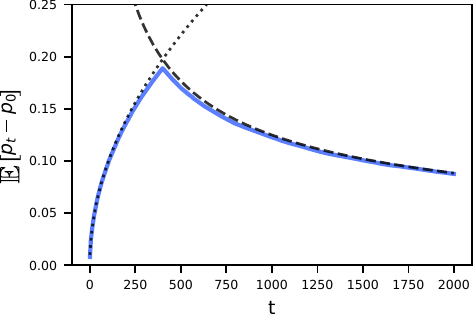}
\caption{Price impact for a buy meta-order ($Y=1$ and $\nu=0.035$) as a function of time. The meta-order extinguishes at time $t=400$. The theoretical decay  $\sim Q/2\sqrt{\pi t}$ (dashed line) matches well with the Monte-Carlo simulation (blue full line).}
\label{fig:decay}
\end{figure}

Note that $\Im(q)/\Im(Q)\simeq\sqrt{Q/q}$ for $q\ge Q$, which is the same behaviour derived in the latent order book model \cite{donier2015fully} and it is compatible with the slow power-law decay predicted by the propagator model \cite{bouchaud2003fluctuations,zarinelli2015beyond} and the empirically observed power-law decay of the impact for intra-day time-scales~\cite{bucci2018slow}.

Our theory predicts a decay of the impact to zero, and no permanent impact (see Fig.~\ref{fig:decay}). 
This is consistent with the fact that, in the original formulation of the GMM, the signal that the informed trader has access to is not permanent either, because she ceases to submit orders after time $T$.

There is no conclusive evidence on whether meta-orders leave a permanent impact or not. Some studies~\cite{brokmann2015slow,gomes2015market} suggest that the decay should tend towards the initial price. Others~\cite{moro2009market,zarinelli2015beyond} support the hypothesis of a plateau. The ``fair pricing'' theory~\cite{farmer2013efficiency} predicts a ``universal'' permanent impact equal to $2/3$ of the peak impact, which seems to be ruled out by more recent empirical studies on larger datasets~\cite{bucci2018slow}. Permanent impact has been related to the effect of informed traders~\cite{bucci2018slow} and it should vanish in perfectly efficient markets, in agreement with the prediction based on the Minority Game~\cite{barato2013impact}. Within our theory, impact decay arises from the interplay between a constant bias in the order flow $n_t$, that leads to $\E{\xi}=GQ/\sqrt{t}$, and a linearly increasing variance of $n_t$, which yields $\V{\xi}\simeq 1$. A permanent impact can arise under an anti-correlated order flow such that the variance of $n_t$ saturates, while the bias in the order flow is preserved. It is easy to see that this would generate a permanent impact which is proportional to $Q$. In the latent order book model~\cite{benzaquen2018market}, a permanent impact arises when the market has a finite memory. In our case, a finite memory can also lead to a constant variance of $n_t$, and hence to $\V{\xi}\sim 1/t$, but at the same time it washes out the bias in the order flow. As a result, no permanent impact arises in our model with a finite memory. These insights were fully confirmed by numerical simulations.


For small $t-T$, our theory predicts that the impact immediately after the end of the meta-order decays as 
\[
\E{\Delta p_{t>T}}/\E{\Delta p_T}\simeq 1-(t-T)/2T\,.
\] 
This is less steep than the decay $\E{\Delta p_{t>T}}/\E{\Delta p_T}\simeq1-\sqrt{(t-T)/T}$
predicted by the latent order books model~\cite{donier2015fully}. 

Our theory also predicts that, upon reverting the direction of the meta-order after a time $T=Q/\nu$, the price should revert to the original price after a time $T$ (see Appendix~\ref{app:reverse}). This contrasts with the latent order books theory, which predicts a faster return to the original price, in a time equal to $T/4$~\cite{donier2015fully}. Numerical simulations confirmed that an asymmetric behaviour with a faster return to the original price attains in our model if a finite memory is introduced.

\subsection{An intuitive argument}

Some insights on the origin of the behaviour derived this far can be gained by a simple argument that assumes that the market maker relies on a Bayesian estimate\footnote{We show in Appendix~\ref{app:infer} that a maximum likelihood approach would lead to a qualitatively different result.} ${\hat\nu_{\rm Bayes}=\E{\nu|x_{\le t}}}$ of $\nu$. This leads to (see Appendix~\ref{app:infer})
\begin{equation}
\label{hatnubayes}
\hat\nu_{\rm Bayes}\simeq \left\{\begin{array}{cc} \bar\nu/2 & t\ll\bar\nu^{-2} \\
\sqrt{\frac{2}{\pi t}} & \bar\nu^{-2}\ll t\ll \nu^{-2}\end{array}\right.
\end{equation}

\begin{figure}[!ht]
\centering
\includegraphics[width=0.8\textwidth]{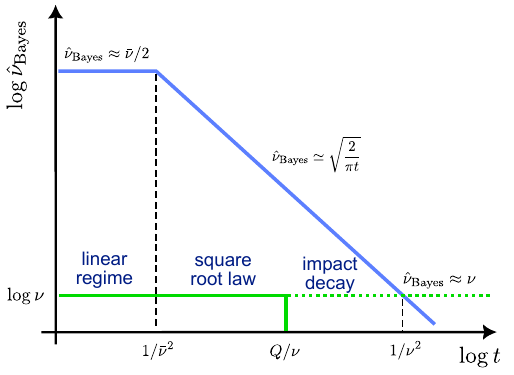}
\caption{Sketch of the behaviour of the Bayesian estimate of $\hat\nu_{\rm Bayes}$ (blue curve). The solid green curve represents the real value of $\nu$ imposed by the informed trader.}
\label{fig:sketch}
\end{figure}

In words, at short times ($t\ll\bar\nu^{-2}$) the prior is dominated by the cutoff $\bar\nu$ whereas for longer time-scales its width shrinks as $1/\sqrt{t}$. Substituting the leading behaviour of $\hat\nu_{\rm Bayes}$ in the linear impact results (Eq.~\ref{linImpact}), i.e. $\Im(q)\simeq \theta\hat\nu_{\rm Bayes} q$, allows one to recover both the SRIL (Eq.~\ref{Epterf}, apart from a factor $\sqrt{2}$) and the crossover to the linear regime (Eq.~\ref{linear}).

Notice that the meta-order execution time $Q/\nu$ is typically much shorter than the time when the true value $\hat\nu_{\rm Bayes}\simeq\nu$ is discovered. Hence, the whole phenomenology of the market impact is concealed by stochastic fluctuations. Indeed, in empirical studies, it  emerges only by carefully taking  averages over many meta-orders, conditional on their inception. 

\section{Conclusion}

In summary, this paper shows that the square-root law of market impact can be derived as an exact result in the relevant scaling limit of a simple model of a market in which the price is formed through a fully Bayesian reasoning. This approach borrows the price formation mechanism from the GMM model, 
combining it with an exogenous term that restores price diffusivity (as e.g. in~\cite{lehalle2019incorporating}). In this framework, we observe that the persistent order flow that originates from a meta-order has the same effect as the activity of an informed trader in the GMM.

The extreme simplicity of the model and the absence of any {\em ad-hoc} assumptions provides further theoretical support for the empirically observed universality of the square root law of market impact. 
The independence of this law on the execution protocol is a consequence of the fact that the scaling regime of interest $t\ll \nu^{-2}$ lies in a region where the contribution from the meta-order to the order flow is statistically undetectable.

This model also allows to recover the crossover to a linear impact for small orders observed in \cite{bucci2019crossover}, as well as the impact decay for times $q\gg Q$. This approach predicts that the price, on average, reverts back to the initial value after the meta-order ends. All this phenomenology can be expressed in terms of the meta-order execution time $q$, and of empirically measurable quantities, in a non-trivial limit of high frequency trading ($\nu\to 0$). 

Besides providing a simple and transparent picture of market impact phenomenology, this theory also predicts that the volatility should acquire a constant term when statistical averages are conditioned on the initiation of a meta-order (Eq.~\ref{variance_mart}). 
A further prediction of our theory (see Appendix~\ref{app:bidask}) is that the average bid-ask spread, conditional on the start of the meta-order, should decrease as $1/\sqrt{t}$. Both predictions could be tested in further empirical studies.

It is important to comment on the limits of our theory, that are an unavoidable consequence of its simplicity, in order to put our results in the correct perspective. As compared to the complexity of real markets, our model makes a number of draconian assumptions. 
First, the model assumes an uncorrelated order flow of unit volume orders of noise traders. We show in Appendix \ref{app:fattail} how our results change under a fat tailed distribution of order volumes of noise traders, showing that the SRIL still holds if the distribution of volumes has a finite second moment, which is consistent with empirical analysis. The SRIL also holds, as shown in Appendix~\ref{app:corr}, for short ranged correlated order flows, but it is modified if order flows exhibits long range auto-correlations. Such long range correlations can also be ascribed to correlated meta-orders. Our theory
describes the response of the market to a single meta-order. Hence it neglects effects induced by correlated meta-orders, 
which have been suggested to be relevant for market impact phenomenology~\cite{bucci2020co}. 

A key assumption in our modelling approach is Eq.~(\ref{sigmatau}). This relates price volatility to the strength of insider information. As a consequence,  we argue that $\nu$ and $\theta$ both depend on the meta-order execution protocol. Yet while in the Bayesian approach $\nu$ is inferred by the market, we implicitly assume that the market maker ``knows'' $\theta$. In a full fledged game theoretic treatment of the interaction between a market maker and an informed trader this would be a clear limit. 

Our approach borrows the price formation mechanism from the GMM as a minimal model of how markets respond to a persistent order flow generated by meta-orders. In doing so, we abstract from strategic considerations on the grounds that meta-order execution is concealed by stochastic fluctuations in the regime of interest. In this perspective, $\theta$ is regarded as a parameter that tunes the strength of market response, that depends on the true value of $\nu$ rather than on its inferred value at time $t$. The independence of our key results Eqs.~(\ref{theoSRIL}) and (\ref{eq:exp_price_decay}) on $\theta$ corroborates the reasonableness of this approach.

In spite of its limits, we believe that the theory presented here provides a coherent stylised model of market impact phenomenology that can be useful in further research. 


 




\section{Acknowledgments}
{We are grateful to Leonardo Bargigli, Jean-Philippe Bouchaud, Fabrizio Lillo, Iacopo Mastromatteo and Michele Vodret for several discussions and stimulating comments.}


\bibliographystyle{unsrt} 
\bibliography{pnas-sample}

\appendix

\section{Linear impact when $\nu$ is known}
\label{app:lin}

Let's recall that the time series $x_{\le t}=(x_1,\ldots,x_t)$ of trades ($x_t=1$ for buy and $x_t=-1$ for sell) 
is a random variable whose distribution $P\{x_{\le t}|G,\nu\}$ is given by Eq.~(\ref{Pxlet}).

Using Bayes rule we find
\[
P\{G=+1|x_{\le t},\nu\} = \frac{P\{x_{\le t}|G=+1,\nu\}P\{G=+1|\nu\}}{P\{x_{\le t}|G=+1,\nu\}P\{G=+1|\nu\}+{P\{x_{\le t}|G=-1,\nu\}P\{G=-1|\nu\}}} \\
\]
and $P\{G=\pm 1|\nu\}=1/2$. This allows us to compute the contribution to the price arising from the order flow, that can be written as
\begin{eqnarray}
\label{EGxnu}
\E{G|x_{\le t},\nu} & = & \frac{(1+\nu)^{n_t}(1-\nu)^{t-n_t}-(1+\nu)^{t-n_t}(1-\nu)^{n_t}}{(1+\nu)^{n_t}(1-\nu)^{t-n_t}+(1+\nu)^{t-n_t}(1-\nu)^{n_t}}\\
 & = & \tanh\left[\left(n_t-\frac{t}{2}\right)\log\frac{1+\nu}{1-\nu}\right]\,.
\end{eqnarray}
Therefore, when the market maker knows $\nu$, the price takes the following explicit expression
\[
p_t(\nu)=F_{t-1}+\theta \tanh\left[\left(n_t-\frac{t}{2}\right)\log\frac{1+\nu}{1-\nu}\right]\,.
\]
Using the fact that $n_t$ is a binomial random variable with mean $\E{n_t}=\frac{1+\nu G}{2}t$ and variance $\V{n_t}=\frac{1-\nu^2}{4}t$, we find Eq.~(\ref{linpt}), to leading order. The leading contribution of informed traders to the expected value of $p_t$ is proportional to $\theta\nu^2 t G$, for $\nu^2 t\ll 1$, whereas its fluctuation is of the order $\theta\nu\sqrt{t}$. 

When the meta-order is over ($t>T$) the statistics of $n_t$ changes as discussed in the main text. The argument of Eq.~(\ref{linpt}) is dominated by the random part, and 
\begin{equation}
\label{EGlin}
\E{G|x_{\le t},\nu} \simeq  \tanh(\nu\sqrt{t}\xi)+ \frac{\nu^2 G T}{\cosh^2(\nu \sqrt{t} \xi)}+O(\nu^4 T^2)
\end{equation}
with $\xi$ a Gaussian random variable with zero mean and unit variance. Taking the expected value over $\xi$, the first term vanishes.
Shortly after the end of the meta-order, i.e. for  $T<t\ll \nu^{-2}$, the second term is constant because $1/\cosh^{2}(\nu \sqrt{t} \xi)\simeq 1-\nu^2 t\xi^2+\ldots$. This leads to an apparent permanent impact as long as $t\ll \nu^{-2}$. Yet, the market impact decays for times $t\gg\nu^{-2}$, because the expected value of the second term in the right hand side of Eq.~(\ref{EGlin}) yields 
\[
\E{\E{G|x_{\le t},\nu}} \simeq \sqrt{\frac{2}{\pi t}} \nu G T\left[1+O(\nu^{-2}t^{-1})\right],\qquad t\gg \nu^{-2}\,.
\]

\section{Derivation of the SRIL}
\label{app:sril}

When $\nu$ is unknown, the contribution to the price arising from the order flow, as shown in the main text, is given by
\begin{equation}
\label{ptnt_app}
\E{G|x_{\le t}}=\int_0^1\!dv \E{G|x_{\le t},v}P(v|x_{\le t})
\end{equation}
where the posterior distribution of $\nu$ is given by
\begin{eqnarray}
P(v|x_{\le t}) = \frac{P(x_{\le t}|v)\phi(v)}{\int_0^1 P(x_{\le t}|y)\phi(y)dy}
\label{bysr}
\end{eqnarray}
and $\phi(v)$ is the prior. Combining this with $\E{G|x_{\le t},v}$ taken from Eq.~(\ref{EGxnu}), Eq.~(\ref{ptnt_app}) becomes
\begin{equation}
\label{eq:ptint}
\E{G|x_{\le t}}=\frac{\int_{-1}^1\!dv\phi(|v|)e^{-tD(z||v)}{\rm sign} v}{\int_{-1}^1\!dv\phi(|v|)e^{-tD(z||v)}}
\end{equation}
where $z=(2n_t-t)/t$ and 
\begin{equation}
\label{KLD}
D(z||x) \equiv \frac{1+z}{2}\log\frac{1+z}{1+x}+\frac{1-z}{2}\log\frac{1-z}{1-x} 
\end{equation}
is a Kullback-Leibler divergence. 

Notice that the random variable $z$ is typically very small for $t\gg 1$. Indeed it can be written as $z=\xi/\sqrt{t}$, where $\xi$ is well approximated by a Gaussian random variable with mean $\E{\xi}=G\nu\sqrt{t}$ and variance $\V{\xi}=1-\nu^2$. Notice also that $\E{\xi}$ is very small in the scaling regime $1\ll t\ll \nu^{-2}$ of interest. 

For $t\to\infty$, the integrals on $v$ in Eq.~(\ref{eq:ptint}) are dominated by values of $v$ such that $D(z||v)\sim t^{-1}$. Since $D(z||v)\sim (z-v)^2$, the integrals are dominated by values of $v$ such that $(v-z)\sim 1/\sqrt{t}$. This implies that the relevant range in the integrals is when $v\sim 1/\sqrt{t}$, because $z\sim 1/\sqrt{t}$. Correspondingly, the integrals probe the behaviour of the prior $\phi(v)$ in the limit $v\to 0$. We shall assume that $\phi(v)$ has a finite limit as $v\to 0$. Different singular behaviours of the prior for $v\to 0$ will be discussed in the next Section.

Summarising, to leading order we find $D(z||v) \simeq \frac{(\gamma-\xi)^2}{2t}+O(1/t^2)$, which leads to
\begin{equation}
\E{G|x_{\le t}} \simeq  \frac{1}{\sqrt{2\pi}}\int_{-\infty}^\infty\!d\gamma  e^{-\frac{1}{2}(\xi-\gamma)^2}{\rm sign}\gamma={\rm erf}\!\left(\xi/\sqrt{2}\right)\,,
 \label{ptxi1}
\end{equation}
which is Eq.~(\ref{ptxi}) in the main text.
Taking the expectation over $\xi$, we find\footnote{Following the same steps, it can be shown that $\E{G|x_{\le t}} \propto \nu\sqrt{t}$ for $t\ll \nu^{-2}$ also if $P\{G=+1\}\neq 1/2$. Indeed, $p_t(\xi)$  can be expressed as a series in odd powers of $\xi$. Considering the leading behavior $\E{\xi^{2n+1}}\simeq \frac{2^{n+1}}{\sqrt{\pi}}\Gamma(n+3/2)\nu\sqrt{t}+O(\nu \sqrt{t})^3$ of the expected values of odd powers of $\xi$, we conclude that the SRIL holds also in this case.} Eq.~(\ref{Epterf}) in the main text.
Notice that, within this approximation,  $\E{G|x_{\le t}}\to G$ as $t\to\infty$.

From Eq.~(\ref{ptxi1}) one can also compute the variance of this contribution to the price
\[
\sqrt{\V{\E{G|x_{\le t}}}} = \frac{1}{\sqrt{3}} - \frac{\sqrt{3}-1}{2\sqrt{3}\pi}\nu^2t+O(\nu t^{1/2})^3\,.
\]

\section{Square-root-law for priors $\phi(\nu){\sim}\nu^{k-1}$ as $\nu\to 0$}
\label{app:prior}

Here we generalize the derivation of the previous Section to priors which behave as $\phi(\nu){\sim}A\nu^{k-1}$ as $\nu\to 0$, with $k>0$ and $A$ a normalization constant to ensure $\int_0^1 \mathrm{d}v \phi(v)=1$. Thus, introducing again $\xi=(2n_t-t)/\sqrt{t}$ and setting $v=\gamma/\sqrt{t}$ in  Eq.~(\ref{eq:ptint}), we have
\begin{equation}
   \E{G|x_{\le t}}\approx\frac{\int_{-\infty}^\infty\!d\gamma |\gamma|^{k-1} e^{-\frac{1}{2}(\xi-\gamma)^2}{\rm sign}\gamma}{\int_{-\infty}^\infty\!d\gamma |\gamma|^{k-1} e^{-\frac{1}{2}(\xi-\gamma)^2}}=\xi\sqrt{2}\frac{\Gamma(\frac{1+k}{2})}{\Gamma(\frac{k}{2})}\frac{_1F_1(1-\frac{k}{2},\frac{3}{2},\frac{-\xi^2}{2})}{_1F_1(\frac{1-k}{2},\frac{1}{2},\frac{-\xi^2}{2})} 
\label{eq:price_prior}
\end{equation}
where ${_1F_1}(a,b,x)$ is the confluent hypergeometric function of parameters $a$ and $b$. The expansion of  Eq.~(\ref{eq:price_prior}) in powers of $\xi$ only contains odd powers. The leading terms are
\[
\E{G|x_{\le t}}= \sqrt{2}\frac{\Gamma\left(\frac{1+k}{2}\right)}{\Gamma\left(\frac{k}{2}\right)}\left[\xi+\frac{1-2k}{6}\xi^3+\frac{(3-4k)(1-4k)}{120}\xi^5 + \dots\right]\,.
\] 
Considering the leading behavior $\E{\xi^{2n+1}}\simeq \frac{2^{n+1}}{\sqrt{\pi}}\Gamma(n+3/2)\nu\sqrt{t}+O(\nu \sqrt{t})^3$ 
 of the expected values of odd powers of $\xi$, we conclude that the SRIL holds also in this case, with a coefficient that depends on $k$.
 For small values of $k$, Eq.~(\ref{eq:price_prior}) reads
\begin{eqnarray}
    \E{G|x_{\le t}} &\simeq& k\frac{\pi}{2}\mathrm{erfi}\left(\frac{\xi}{\sqrt{2}}\right)+O(k^2) \\
    &\simeq& k\sqrt{\frac{\pi}{2}}\left[\xi+\frac{\xi^3}{6}+\frac{\xi^5}{40}+O(\xi^7)\right] \nonumber
\end{eqnarray}
where $\mathrm{erfi}(x)=\frac{2}{\sqrt{\pi}}\int_0^x e^{z^2} dz$ is the imaginary error function. In the limit of small $k$, the market impact varies linearly with $k$ and vanishes for $k\to0$. In fact for $k=0$, the prior behaves as a delta-function at $\nu=0$, that corresponds to a market maker who believes that there is no informed trader (or meta-order). Yet, at time $t\gg\nu^{-2}$ the true value of $\nu$ is revealed irrespective of what the prior is. Therefore, the behavior $\E{\Delta p_t}\propto \theta k\nu\sqrt{t}$ at $t\ll\nu^{-2}$ should leave way to a faster growth of the market impact, as shown in Fig.~1 of the main paper, 
in the $t\sim\nu^{-2}$ regime.

\section{Square-root law for generic order flow processes}
\label{app:vol}

In the main text, we considered that all agents trade a unit volume at each step of the game, and that orders generated by noise traders are uncorrelated. We can generalize the model by allowing both the informed trader and the noise traders to trade non unitary volumes at the same time, and by allowing correlated order flows. In order to be general, we will also assume that $G$ is worth $1$ with probability $p$ and $-1$ with probability $1-p$. 

At each step $t$ of the game, the informed trader will buy a quantity $\chi>0$ of the asset if $G=1$ and she will sell a volume $-\chi<0$ if $G=-1$ (so that $\chi$ is always positive).

Noise traders trade randomly. At each step $t$, they trade a volume $v_t$ which follows a normal law $\mathcal{N}(r,\sigma_v^2)$, where $r$ quantifies the bias of the noise traders and $\sigma_v$ is the typical volume of uninformed trades. To simplify the analysis, we will study the case of unbiased noise traders so $r=0$. 

The market maker is the one who provides the liquidity of the market. At each step he fixes the price at which he sells or buys the asset in order to make zero profit on average. Hence the price is $p_t=F_{t-1}+\theta\E{G|x_{\le t}}$. The only information he has is the volume imbalance $\Delta V$ at time $t$ and the probability $p$ of $G$ to be equal to $1$. Let's start the analysis by assuming that the market maker knows the speed of trade of the informed trader $\chi$. 

Let us introduce $\mathcal{V}_t=\sum_{\tau=1}^t v_t$ the cumulative volume traded by noise traders until time $t$. Thus $\mathcal{V}_t$ follows the Gaussian distribution $\mathcal{N}(0,\sigma_v^2t)$. The volume imbalance at time $t$ is then $\Delta V=\mathcal{V}_t + G\chi t$, which is also a Gaussian variable with distribution $P(\Delta V|G,\chi)\propto\exp\left(-\frac{1}{2}\frac{(\Delta V -G \chi t)^2}{\sigma_v t^2}\right)$. Therefore,
\begin{eqnarray}
    \E{G|\Delta V,\chi} &=& P(G=1|\Delta V,\chi)-P(G=-1|\Delta V,\chi) \nonumber \\
    &=& \frac{pP(\Delta V|G=1,\chi)-(1-p)P(\Delta V|G=-1,\chi)}{pP(\Delta V|G=1,\chi)+(1-p)P(\Delta V|G=-1,\chi)} \nonumber \\
    &=& \frac{p\exp\left(-\frac{1}{2}\frac{(\Delta V-\chi t)^2}{\sigma_v^2t}\right)-(1-p)\exp\left(-\frac{1}{2}\frac{(\Delta V+\chi t)^2}{\sigma_v^2t}\right)}{p\exp\left(-\frac{1}{2}\frac{(\Delta V-\chi t)^2}{\sigma_v^2t}\right)+(1-p)\exp\left(-\frac{1}{2}\frac{(\Delta V+\chi t)^2}{\sigma_v^2t}\right)} \nonumber \\
    &=& \left[1-\frac{1-p}{p}\exp\left(-\frac{2\chi\Delta V}{\sigma_v^2}\right)\right]/\left[1+\frac{1-p}{p}\exp\left(-\frac{2\chi\Delta V}{\sigma_v^2}\right)\right] \label{eq:q_known} \,.
\end{eqnarray}

Let's now consider the case where the speed of trade $\chi$ is not known. Thus we have
\begin{equation}
    \E{G|\Delta V}=\int_0^\infty d\chi \E{G|\Delta V,\chi}P(\chi|\Delta V) \,. \label{eq:price_q}
\end{equation}

Knowing $\Delta V$, the probability distribution of $\chi$ is
\begin{eqnarray}
    P(\chi|\Delta V) &=& \frac{P(\Delta V|\chi)\phi(\chi)}{\int_0^\infty d\chi P(\Delta V|\chi)\phi(\chi)} \nonumber \\
    &=& \frac{\left[pP(\Delta V|G=1,\chi)+(1-p)P(\Delta V|G=-1,\chi)\right]\phi(\chi)}{\int_0^\infty d\chi [pP(\Delta V|G=1,\chi)+(1-p)P(\Delta V|G=-1,\chi )]\phi(\chi)} \label{eq:prob_q}
\end{eqnarray}
where $\phi(\chi )$ is the prior distribution of the speed of trade $\chi $.
By injecting Eq.~(\ref{eq:q_known}) and Eq.~(\ref{eq:prob_q}) in Eq.~(\ref{eq:price_q}) we get
\begin{equation}
    \E{G|\Delta V}=2\left[1+\frac{1-p}{p}\frac{\int_0^\infty d\chi \phi(\chi )\exp\left(-\frac{1}{2}\frac{(\Delta V+\chi t)^2}{\sigma_v^2t}\right)}{\int_{-\infty}^0 d\chi \phi(\chi )\exp\left(-\frac{1}{2}\frac{(\Delta V+\chi t)^2}{\sigma_v^2t}\right)}\right]^{-1}-1 \,.
\end{equation}

If we take an uniform prior, we have
\begin{equation}
    \E{G|\Delta V}=\left[1-\frac{1-p}{p}\frac{1-\mathrm{erf}\left(\frac{\Delta V}{\sqrt{2t}\sigma_v}\right)}{1+\mathrm{erf}\left(\frac{\Delta V}{\sqrt{2t}\sigma_v}\right)}\right]/\left[1+\frac{1-p}{p}\frac{1-\mathrm{erf}\left(\frac{\Delta V}{\sqrt{2t}\sigma_v}\right)}{1+\mathrm{erf}\left(\frac{\Delta V}{\sqrt{2t}\sigma_v}\right)}\right] \label{eq:price_gen}
\end{equation}
and for the special case $p=\frac{1}{2}$, this equation becomes
\begin{equation}
    \E{G|\Delta V}=\mathrm{erf}\left(\frac{\Delta V}{\sqrt{2t}\sigma_v}\right) \label{eq:price_1/2}
\end{equation}
which is very similar to the one obtained in the unit volume trading model (Eq.~\ref{ptxi1}). Here the analogous of $\xi$ is $\Delta V/(\sigma_v\sqrt{t})\sim\mathcal{N}\left(G\frac{\chi }{\sigma_v}\sqrt{t},1\right)$. It is also easy to see that the quantity analogous to $\nu$ is $\chi /\sigma_v$. By taking the expectation over the transactions of noise traders, we obtain the price impact
\begin{equation}
    \E{\Delta p_t(\Delta V)}=\theta\E{\E{G|\Delta V}}=G \theta\mathrm{erf}\left(\frac{\chi \sqrt{t}}{2\sigma_v}\right)= G \theta\frac{\chi \sqrt{t}}{\sqrt{\pi}\sigma_v}+O\left(\frac{\chi \sqrt{t}}{\sigma_v}\right)^2
    \label{eq:impact_gene}
\end{equation}
which again features the square root behavior. From Eq.~(\ref{eq:price_1/2}), one can also compute the standard deviation of the order imbalance contribution
\begin{equation}
\label{sigma_nonunit}
\sigma=\sqrt{\V{\E{G|\Delta V}}} = \frac{1}{\sqrt{3}} - \frac{\sqrt{3}-1}{2\sqrt{3}\pi}\left(\frac{\chi }{\sigma_v}\right)^2t+O\left(\frac{\chi }{\sigma_v} t^{1/2}\right)^3\,.
\end{equation}

\subsection{Response of the market to order flow imbalance}

Thanks to Eq.~(\ref{eq:price_gen}) and Eq.~(\ref{eq:price_1/2}), we can compute the price as a function of the order flow imbalance $\Delta V$ at time $t$. Thus, the aggregated impact $\E{\Delta p_t|\Delta V}$ of the market can be computed from these expressions as the martingale contribution will be washed by the expectation. 


An interesting parameter that can be extracted from the aggregated impact is the Kyle lambda $\Lambda$. It is the slope of the aggregated impact at $\Delta V=0$ and it quantifies the response of the market to a small order flow imbalance. By taking the derivative of Eq.~(\ref{eq:price_gen}) at $\Delta V=0$ and injecting it into $p_t=F_{t-1}+\theta\E{G|\Delta V}$, we obtain
\begin{equation}
    \Lambda=4\theta\sqrt{\frac{2}{\pi t}}\frac{p(1-p)}{\sigma_v}
\end{equation}
so the Kyle lambda in our model decreases as a function of time, as observed empirically \cite{bouchaud2018trades}. This contrasts with the prediction of the Kyle model where $\Lambda$ is a constant~\cite{kyle1985continuous}. Notice also that the response of the market increases with $\V{G}=4p(1-p)$, i.e. with the uncertainty on the value of $G$. 

By noticing that the Bayesian estimator of the speed of trade of the informed trader reads
$\chi^\mathrm{Bayes}=\int_0^\infty \mathrm{d}\chi ~\chi P(\chi |\Delta V)=\sqrt{\frac{2}{\pi t}}\sigma_v+O(\Delta V)$, one can recast the Kyle lambda as
\begin{equation}
\label{KyleBayes}
    \Lambda=\theta \chi^\mathrm{Bayes}\frac{\V{G}}{\sigma_v^2} \,,
\end{equation}
emphasizing the fact that the response of the market maker to the volume imbalance is proportional to his estimate of the speed of trade of the informed trader. This is not surprising because if we expand Eq.~(\ref{eq:q_known}), which describes the response of the market maker when $\chi $ is known, we find
\begin{equation}
    \E{G|\Delta V}=(2p-1)+\theta \chi \frac{\V{G}}{\sigma_v^2}\Delta V+O(\Delta V^2)\,.
\end{equation}
Eq.~(\ref{KyleBayes}) is consistent with this equation where the value of $\chi $ is replaced by its Bayes estimator. 

\subsection{Fat tailed distribution for the noise trading volumes}
\label{app:fattail}

The derivation of the previous Section can be generalised to the case where noise traders submit orders with a fat tailed volume distribution. Let's assume that the volume $v_t$ traded by noise traders at time $t$ follows an even distribution with $P(v)\sim\frac{C}{|v|^{1+\alpha}}$ for $v\to\infty$, where $C$ is a constant and $\alpha$ is the exponent of the distribution. For $\alpha\ge2$, the variance $\sigma_v$ of $v_t$ is finite. Therefore, the central limit theorem (CLT) applies and the cumulative volume $\mathcal{V}_t$ of noise traders follows the normal law distribution $\mathcal{N}(0,\sigma_v^2t)$. In summary, the behavior of $\Delta V$ is asymptotically the same than for Gaussian orders of noise traders when $t\to\infty$. Thus we recover the square root law for all $\alpha\ge 2$. 

For $\alpha<2$, we can not apply the CLT because the variance of $v_t$ is not finite. Here, we will assume that the $v_t$ follow Levy stable distribution $L_{\alpha,\sigma_v}$, with $\alpha$ the stability parameter, $\sigma_v$ the scale parameter and both the location parameter and skewness parameter equal to zero in order to have an even distribution. This distribution can be seen as a generalisation of the Gaussian distribution which we recover for $\alpha=2$. Also, this distribution has the asymptotic behaviour $L_{\alpha,\sigma_v}(x)\sim\frac{C(\alpha)}{|x|^{1+\alpha}}$. It has the nice property that, if ${v}_\tau$ are independent random variables with distribution $L_{\alpha,\sigma_v}$ for all $\tau$, then $\mathcal{V}_t=\sum_{\tau=1}^t v_\tau$ follows the $L_{\alpha,\sigma_v t^\frac{1}{\alpha}}$ Levy distribution.


Thanks to these properties, and by injecting $\mathcal{V}_t$ in Eq.~(\ref{eq:q_known}) to (\ref{eq:prob_q}) (remember that $\Delta V=\mathcal{V}_t +G\chi t$), for $p=1/2$ we obtain
\begin{equation}
    \E{G|\Delta V}=2\int_{-\infty}^\frac{\Delta V}{\sigma_v t^{1/\alpha}} L_{\alpha,1}(x)dx-1 \,.
\end{equation}

Introducing $\mathcal{L}_{a,c}(x)=\int_0^x L_{a,c}(x)dx$, as $\lim\limits_{x\to-\infty}\mathcal{L}_{a,c}(x)=-\frac{1}{2}$, we can rewrite the last expression
\begin{equation}
    \E{G|\Delta V}=2\mathcal{L}_{\alpha,1}\left(\frac{\Delta V}{\sigma_v t^{1/\alpha}}\right) \,.
\end{equation}

We can give some examples of Alpha-L\'evy distributions. For $\alpha=2$ we recover the Gaussian distribution and $\mathcal{L}_{2,1}(x)=\frac{1}{2}\mathrm{erf}(x)$. For $\alpha=1$, we have the Cauchy distribution and $\mathcal{L}_{1,1}(x)=\frac{1}{\pi}\arctan(x)$.

Let us now compute the price impact. If we consider $\chi t\ll \mathcal\sigma_v t^{1/\alpha}$, we will have
\begin{eqnarray}
    \mathcal{L}_{\alpha,1}\left(\frac{\Delta V}{\sigma_v t^{1/\alpha}}\right) &=& \mathcal{L}_{\alpha,1}\left(\frac{\mathcal{V}_t+\chi t}{\sigma_v t^{1/\alpha}}\right) \nonumber \\
    &=& \mathcal{L}_{\alpha,1}\left(\frac{\mathcal{V}_t}{\sigma_v t^{1/\alpha}}\right)+L_{\alpha,1}\left(\frac{\mathcal{V}_t}{\sigma_v t^{1/\alpha}}\right)\frac{\chi t}{\sigma_v t^{1/\alpha}}+O\left(\frac{\chi }{\sigma_v}t^{1-\frac{1}{\alpha}}\right)^2
\end{eqnarray}

So the expected price reads
\begin{eqnarray}
    \E{p_t} &=& 2\theta\int_\mathbb{R} du~L_{\alpha,\sigma_v t^{1/\alpha}}(u)\mathcal{L}_{\alpha,1}\left(\frac{u+\chi t}{\sigma_v t^{1/\alpha}}\right) \nonumber \\
    &=& 2\theta\int_\mathbb{R} du~L_{\alpha,\sigma_v t^{1/\alpha}}(u)\mathcal{L}_{\alpha,1}\left(\frac{u}{\sigma_v t^{1/\alpha}}\right) \\
    & ~ & + 2\theta\frac{\chi t}{\sigma_v t^{1/\alpha}}\int_\mathbb{R} du~L_{\alpha,\sigma_v t^{1/\alpha}}(u)L_{\alpha,1}\left(\frac{u}{\sigma_v t^{1/\alpha}}\right)+O\left(\frac{\chi }{\sigma_v}t^{1-\frac{1}{\alpha}}\right)^2 \nonumber \\
    &=& 0+2\theta K(\alpha)\frac{\chi t^{1-\frac{1}{\alpha}}}{\sigma_v}+O\left(\frac{\chi }{\sigma_v}t^{1-\frac{1}{\alpha}}\right)^2
\end{eqnarray}
where $K(\alpha)=\int_\mathbb{R} du~L_{\alpha,1}(u)^2$. For $\alpha=2$, $K(2)=1/(2\sqrt{\pi})$ so we recover Eq.~(\ref{eq:impact_gene}).

In financial market, the exponent of the fat tail is generally reported to be $\alpha\approx 5/2>2$ \cite{bouchaud2018trades} but this exponent can vary from study to study \cite{lillo2005theory,gopikrishnan2000statistical,farmer2004origin}.

\subsection{Square-root law for correlated order flow}
\label{app:corr}

In order to add correlations between the orders, let us assume that the sequence $(v_1,\dots,v_t)$ follows a multivariate normal distribution $\mathcal{N}(\mathbf{0},\mathbf{C})$. $\mathbf{0}$ is the null vector, it indicates that the mean of each $v_t$ is zero. $\mathbf{C}$ is the correlation matrix defined by $(\mathbf{C})_{t,t'}=\E{v_t v_{t'}}=f(|t-t'|)$, where $f(0)=\sigma_v^2$ is the variance of $v_t$ and $f$ is a decreasing function. Thus, the cumulative noise volume $\mathcal{V}_t$ follows the normal distribution $\mathcal{N}(0,\Sigma_t^2)$ where $\Sigma_t^2=\sum_{t'=1}^t\sum_{t''=1}^t f(|t''-t'|)$.

Now, the equivalent of Eq.~(\ref{eq:price_1/2}) is
\begin{equation}
    \E{G|\Delta V}=\mathrm{erf}\left(\frac{\Delta V}{\sqrt{2}\Sigma_t}\right)
\end{equation}
which leads to
\begin{equation}
    \E{\Delta p_t(\Delta V)}=\theta\,\mathrm{erf}\left(\frac{\chi t}{2\Sigma_t}\right) \,.
\end{equation}

Let's consider the case where we have a fast decay of the order correlations. For example we can take $(\mathbf{C})_{t,t'}=\sigma_v^2 \exp\left(-\frac{|t-t'|}{\tau_c}\right)$ where $\tau_c$ is a correlation time. We can now compute the variance of $\mathcal{V}_t$
\begin{eqnarray}
    \Sigma_t^2 &=& t\sigma_v^2+2\sigma_v^2\sum_{1\le t'<t''\le t}\exp\left(-\frac{t''-t'}{\tau_c}\right) \nonumber \\
    &\sim& \left(1+\frac{2}{e^{1/\tau_c}-1}\right)\sigma_v^2 t \,.
\end{eqnarray}

\begin{figure}
\centering
\includegraphics[width=0.8\textwidth]{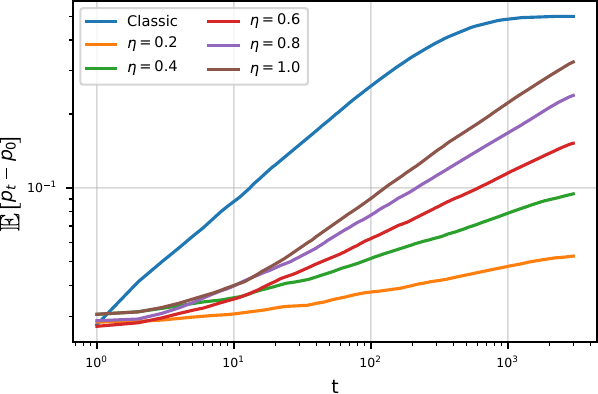}
\caption{Numerical simulation of the price impact for correlated noise trades in the non unit volume model ($\chi =0.1$ and $\sigma_v=1$). The blue curve shows the price impact in the case of uncorrelated noise trades.}
\label{fig:correl}
\end{figure}

For $\tau_c\to0$, $\Sigma_t^2=\sigma_v^2 t$ so we recover the case where the $v_t$ are i.i.d. random variables. One can also remark that the correlated case does not differ so much from the independent case as they both have a linear time dependence. Hence, the square root impact law should hold in the case of short range auto-correlations in the orders of noise traders.

Let's now consider the case where we have long range correlations between market orders, by choosing \footnote{We use this particular correlation matrix because its Fourier transform can be expressed analytically, which allows to generate quality samples for numerical simulations (see \cite{makse1996method}).}
\begin{equation}
    (\mathbf{C})_{t,t+\tau}=\frac{\sigma_v^2}{(1+\tau^2)^\frac{\eta}{2}}{\sim}\frac{\sigma_v^2}{\tau^\eta}~~~({\tau \to +\infty})\,.
\end{equation}

For $\eta=1$, one can show that $\Sigma_t\sim\sigma_v\sqrt{2t\log(t)}$ and for $0<\eta<1$, $\Sigma_t\sim\sqrt{\frac{2}{(1-\eta)(2-\eta)}}\sigma_vt^{1-\frac{\eta}{2}}$. Thus for $0<\eta<1$, we get a price impact law
\begin{equation}
    \E{p_t}\simeq\theta\sqrt{\frac{(1-\eta)(2-\eta)}{2\pi}}\frac{\chi }{\sigma_v}t^{\eta/2}+O\left(\frac{\chi }{\sigma_v}t^{\eta/2}\right)^2
\end{equation}
with a different exponent with respect to the SRIL.
The price impact is plotted for several value of $\eta$ in Fig.~\ref{fig:correl}.

\section{Crossover to linear behaviour}
\label{app:cross}

It makes sense to assume that the market anticipates that the contribution of meta-orders to the order flow is small, i.e. that $\nu\ll \bar\nu\ll 1$. One way to incorporate this observation is to take $\phi(y)=1/\bar \nu$ for $y\in[0,\bar \nu]$ and $\phi(y)=0$ otherwise. With this assumption, Eq.~(\ref{eq:ptint}) becomes
\begin{equation}
\E{G|x_{\le t}}=\frac{\int_{-\bar\nu}^{\bar\nu}\!dv~e^{-tD(z||v)} {\rm sign}v}{\int_{-\bar\nu}^{\bar\nu}\!dv~e^{-tD(z||v)}} \,.
\end{equation}

Setting $v=\gamma/\sqrt{t}$ and introducing as before $\xi=z\sqrt{t}=(2n_t-t)/\sqrt{t}$, we obtain 
\begin{equation}
    \E{G|x_{\le t}} \simeq \frac{\int_{-\bar\nu\sqrt{t}}^{\bar\nu\sqrt{t}}\mathrm{d}\gamma~e^{-\frac{1}{2}(\gamma-\xi)^2}{\rm sign}\gamma}{\int_{-\bar\nu\sqrt{t}}^{\bar\nu\sqrt{t}}\mathrm{d}\gamma~e^{-\frac{1}{2}(\gamma-\xi)^2}}
    \simeq 2\frac{{\rm erf}\!\left(\frac{\xi}{\sqrt{2}}\right)+{\rm erf}\!\left(\frac{\bar\nu\sqrt{t}-\xi}{\sqrt{2}}\right)}{{\rm erf}\!\left(\frac{\bar\nu\sqrt{t}+\xi}{\sqrt{2}}\right)+{\rm erf}\!\left(\frac{\bar\nu\sqrt{t}-\xi}{\sqrt{2}}\right)}-1 \,.
\end{equation}

For $\bar\nu\sqrt{t}\gg 1$ (i.e. $t\gg\bar\nu^{-2}$), we recover $\E{G|x_{\le t}} \simeq {\rm erf}\!\left(\xi/\sqrt{2}\right)$ whereas for $t\ll\bar\nu^{-2}$ the expansion of the expression above for $\bar\nu\sqrt{t}\ll 1$ leads to $\E{G|x_{\le t}}\simeq \frac 1 2 \xi\bar\nu\sqrt{t}+O(\bar\nu^2 t)$. Taking the expectation 
on $\xi$ we find a linear regime, Eq.~(\ref{linear}) in the main text.

\section{Impact decay upon reversing the direction of the meta-order}
\label{app:reverse}

In this section, we will be interested in the case where a meta-order is executed until time $T=Q/\nu$ and is immediately followed by another meta-order in the opposite direction. The statistics of $\xi$ is easily calculated taking into account that 
\begin{equation}
\E{n_t}=\frac{t}{2}+\frac{1}{2}\sum_{\tau=1}^{Q/\nu}\E{x_\tau}+\frac{1}{2}\sum_{\tau=Q/\nu+1}^t\E{x_\tau}=\frac t 2 +G \left(Q-\frac{\nu t}{2}\right)
\label{eq:sum_decay}
\end{equation}
because $\E{x_\tau}=G\nu$ in the first sum and $\E{x_\tau}=-G\nu$ in the second. Similarly we compute the variance $\V{n_t}=\frac{1-\nu^2}{4}t$. Hence $\xi$ is asymptotically a Gaussian variable with mean $\E{\xi}=G\frac{2Q-\nu t}{\sqrt{t}}$ and variance $\V{\xi}=1-\nu^2$. Using this in Eq.~(\ref{ptxi1}), one gets 
\begin{equation}
\E{\Delta p_t}=\E{\theta \E{G|x_{\le t}}}\simeq G \theta \,{\rm erf}\!\left(\frac{Q}{\sqrt{t}}-\frac{\nu\sqrt{t}}{2}\right) \,.
\end{equation}

From this last equation, one can see that the time needed to go back to the original price after the end of the first meta-order is $t-T=T$. Note that for large values of $t$, we recover the square root impact law but in the opposite direction. For small $t-T$ instead, the impact immediately after the end of the first meta-order decays as $\E{p_{t>T}}/\E{p_T}\simeq 1-\frac{3}{2}\frac{t-T}{T}$. This decay is steeper than the case where no meta-order is executed after the first one. In the latter case we find $\E{p_{t>T}}/\E{p_T}\simeq 1-\frac{1}{2}\frac{t-T}{T}$.

\section{Inference of $\nu$}
\label{app:infer}

\begin{figure}
\centering
\includegraphics[width=0.8\textwidth]{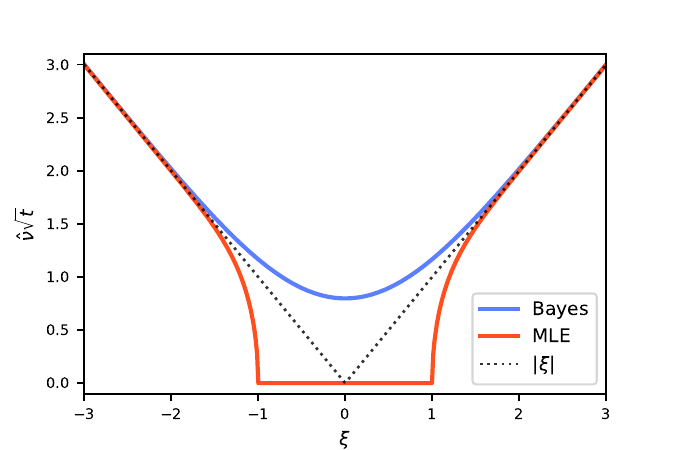}
\caption{Plot of estimators of $\nu$ times $\sqrt{t}$ as a function of $\xi=(2n_t-t)/\sqrt{t}$.}
\label{fig:bayes_mle}
\end{figure}

This section details the derivation of the Bayes estimator and of the maximum likelihood estimator (MLE) of $\nu$ discussed in the main text. Their behaviour is shown in Fig.~\ref{fig:bayes_mle}.

\subsection{Bayesian estimator}

The Bayes estimator is
\begin{equation}
\label{byse}
\hat\nu_\mathrm{Bayes}=\mathbb{E}[\nu|x_{\le t}]=\int_0^1 v P(v|x_{\le t})\mathrm{d}v \,.
\end{equation}

By injecting Eq.~(\ref{bysr}) into Eq.~(\ref{byse}), one can obtain
\begin{equation}
\label{eq:nubayes_exa}
\hat\nu_\mathrm{Bayes}=\frac{\int_{-1}^1\!dv\phi(|v|)|v|e^{-tD(z||v)}}{\int_{-1}^1\!dv\phi(|v|)e^{-tD(z||v)}}
\end{equation}

Again, the asymptotic behaviour for $t\to\infty$ is derived with the change of variables $\xi=z\sqrt{t}$ and $v=\gamma/\sqrt{t}$. 
By assuming that $\phi(v)$ is finite as $v\to0$, we get to the first order in the scaling regime $1\ll t\ll \nu^{-2}$
\begin{eqnarray}
\hat\nu_\mathrm{Bayes} & \simeq&  \frac{1}{\sqrt{2\pi t}}\int_{-\infty}^\infty |\gamma| e^{-\frac{1}{2}(\xi-\gamma)^2}\mathrm{d}\gamma \nonumber\\
        &=& \sqrt{\frac{2}{\pi t}}e^{-\frac{1}{2}\xi^2}+\frac{\xi}{\sqrt{t}}{\rm erf}\!\left(\xi/\sqrt{2}\right) \label{eq:nubayes} 
\end{eqnarray}
which has the limiting behaviours $\hat\nu_\mathrm{Bayes}\simeq \frac{2+\xi^2}{\sqrt{2\pi t}}$ for $|\xi|\ll 1$ and $\hat\nu_\mathrm{Bayes}\simeq \frac{|\xi|}{\sqrt{t}}$ for $|\xi|\gg 1$.

The expected value of $\hat\nu_\mathrm{Bayes}$ over the distribution of $\xi$ is
\[
\E{\hat\nu_\mathrm{Bayes}}=\frac{2}{\sqrt{\pi t}}e^{-\frac{-\nu^2 t}{4}}+\nu~\mathrm{erf}\left(\frac{\nu\sqrt{t}}{2}\right) \,.
\]
Its limiting behaviour is $\E{\hat\nu_\mathrm{Bayes}}\simeq \frac{2}{\sqrt{\pi t}}$ for $t\ll 1/\nu^2$ and $\E{\hat\nu_\mathrm{Bayes}}\propto\nu$ for $t \sim 1/\nu^2$.

\begin{figure}
\centering
\includegraphics[width=0.8\textwidth]{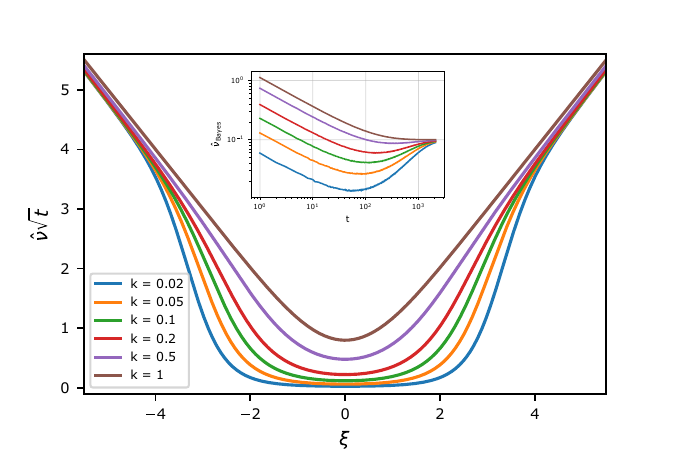}
\caption{Plot of the Bayes estimator times $\sqrt{t}$ as a function of $\xi$ for priors of type $\phi(v)\sim A v^{k-1}$. In the inset, numerical simulations ($\nu=0.1$) of the expected value of the Bayes estimator as a function of time.}
\label{fig:bayes_prior}
\end{figure}

{For a prior of type $\phi(v)\sim A v^{k-1}$,} the same reasoning as above applies and one can show that
\begin{equation}
    \hat\nu_\mathrm{Bayes} \simeq \sqrt{\frac{2}{t}}\frac{\Gamma \left(\frac{k+1}{2}\right)}{\Gamma \left(\frac{k}{2}\right)}\frac{_1F_1\left(-\frac{k}{2},\frac{1}{2},-\frac{\xi^2}{2}\right)}{_1F_1\left(\frac{1-k}{2},\frac{1}{2},-\frac{\xi^2}{2}\right)} 
\end{equation}
For $k=1$, one can recover Eq.~(\ref{eq:nubayes}). The behavior of $\hat\nu_\mathrm{Bayes}$ is plotted for several values of $k$ in Fig.~\ref{fig:bayes_prior}. 
\\

In the case where the prior has a cutoff, i.e.  $\phi(y)=1/\bar \nu$ for $y\in[0,\bar \nu]$ and $\phi(y)=0$ otherwise, Eq.~(\ref{eq:nubayes_exa}) becomes
\begin{eqnarray}
    \hat\nu_\mathrm{Bayes} &=& \frac{\int_{-\bar\nu}^{\bar\nu}\!dv\phi(|v|)|v|e^{-tD(z||v)}}{\int_{-\bar\nu}^{\bar\nu}\!dv\phi(|v|)e^{-tD(z||v)}} \nonumber \\
    &\simeq& \frac{1}{\sqrt{t}}\frac{\int_{-\bar\nu\sqrt{t}}^{\bar\nu\sqrt{t}}\mathrm{d}\gamma~|\gamma|e^{-\frac{1}{2}(\gamma-\xi)^2}}{\int_{-\bar\nu\sqrt{t}}^{\bar\nu\sqrt{t}}\mathrm{d}\gamma~e^{-\frac{1}{2}(\gamma-\xi)^2}} \nonumber \\
    &\simeq& \frac{\sqrt{\frac{2}{\pi t}}}{{\rm erf}\!\left(\frac{\bar\nu\sqrt{t}+\xi}{\sqrt{2}}\right)+{\rm erf}\!\left(\frac{\bar\nu\sqrt{t}-\xi}{\sqrt{2}}\right)}\left[
    2e^{-\frac{\xi^2}{2}}-e^{-\frac{1}{2}(\xi-\bar\nu\sqrt{t})^2}-e^{-\frac{1}{2}(\xi+\bar\nu\sqrt{t})^2}\right.\nonumber\\
    &~&~~~~~~~~~~~~~~\left.+\sqrt{2\pi}\xi{\rm erf}\!\left(\frac{\xi}{\sqrt{2}}\right)-\sqrt{\frac{\pi}{2}}\xi{\rm erf}\!\left(\frac{\xi-\bar\nu\sqrt{t}}{\sqrt{2}}\right)-\sqrt{\frac{\pi}{2}}\xi{\rm erf}\!\left(\frac{\xi+\bar\nu\sqrt{t}}{\sqrt{2}}\right)\right]\nonumber
\end{eqnarray}

Note that we recover Eq.~(\ref{eq:nubayes}) in the limit $\bar\nu\sqrt{t}\gg 1$. However for $\bar\nu\sqrt{t}\ll 1$, the expansion of the expression leads to $\hat\nu_\mathrm{Bayes} \simeq \bar\nu/2+\frac{1}{\sqrt{t}}O(\bar\nu\sqrt{t})^2$.

\subsection{Maximum likelihood estimator (MLE)}

The maximum likelihood estimator is
\[
\nu_{\mathrm{MLE}}=\arg \max_v P(x_{\le t}|v) \,.
\]
We have for $P(G=\pm1)=\frac{1}{2}$
\begin{eqnarray}
P(x_{\le t}|v) & \propto & (1+v)^{n_t}(1-v)^{t-n_t}+(1+v)^{t-n_t}(1-v)^{n_t} \nonumber\\
    &\propto& (1-v^2)^\frac{t}{2}\cosh\left[\frac{\xi\sqrt{t}}{2}\log\left(\frac{1+v}{1-v}\right)\right] \,.
\end{eqnarray}
Hence the condition for the maximum likelihood reads
\[
\frac{\partial P(x_{\le t}|\nu)}{\partial \nu}=0 \Leftrightarrow \nu\sqrt{t}\cosh\left[\frac{\xi\sqrt{t}}{2}\log\left(\frac{1+\nu}{1-\nu}\right)\right]=\xi \sinh\left[\frac{\xi\sqrt{t}}{2}\log\left(\frac{1+\nu}{1-\nu}\right)\right]\,.
\]
Therefore the MLE satisfies the self consistent equation
\begin{equation}
\label{}
\hat\nu_{\mathrm{MLE}}=\frac{\xi}{\sqrt{t}}\tanh\left[\frac{\xi\sqrt{t}}{2}\log\left(\frac{1+\hat\nu_{\mathrm{MLE}}}{1-\hat\nu_{\mathrm{MLE}}}\right)\right] \,.
\end{equation}
For $\xi<1$, this equation has only zero as a solution. For $\xi>1$, the zero solution becomes unstable and $\hat\nu_{\mathrm{MLE}}\simeq \frac{|\xi|}{\sqrt{t}}$ for $|\xi|\gg \sqrt{t}$.

\section{Bid-ask spread}
\label{app:bidask}

The ask and the bid prices at time $t+1$ are given by
\begin{eqnarray}
    a_{t+1} &=& F_t+\theta \int_0^1\!dv \E{G|x_{\le t},x_{t+1}=1,v}P(v|x_{\le t},x_{t+1}=1)\label{askeq}\\
    b_{t+1} &=& F_t+\theta \int_0^1\!dv \E{G|x_{\le t},x_{t+1}=0,v}P(v|x_{\le t},x_{t+1}=0)\label{bideq}
\end{eqnarray}
which can be recast, using Bayes rule, as 
\begin{eqnarray}
    a_{t+1}(z) & = & F_t+\theta \frac{\int_{-1}^1\!dv\phi(|v|)(1+v)e^{-tD(z||v)}{\rm sign}v}{\int_{-1}^1\!dv\phi(|v|)(1+v)e^{-tD(z||v)}}\\
    b_{t+1}(z) & = & F_t+\theta \frac{\int_{-1}^1\!dv\phi(|v|)(1-v)e^{-tD(z||v)}{\rm sign}v}{\int_{-1}^1\!dv\phi(|v|)(1-v)e^{-tD(z||v)}}
\end{eqnarray}
where $z=(2n_t-t)/t$ and $D(z||x)$ is a Kullback-Leibler divergence.  Again, $z$ is typically of order $1/\sqrt{t}$, so we focus on the variable $\xi=z\sqrt{t}$ and make the change of variable $\gamma=v\sqrt{t}$. By also assuming that $\phi(\nu)$ is finite as $\nu\to0$, we get to the first order in the scaling regime $1\ll t\ll \nu^{-2}$
\begin{equation}
    \frac{\int_{-1}^1\!dv\phi(v)(1\pm v)e^{-tD(z||v)}{\rm sign}v}{\int_{-1}^1\!dv\phi(|v|)(1\pm v)e^{-tD(z||v)}}\simeq\frac{\int_{-\infty}^\infty \left(1\pm\frac{\gamma}{\sqrt{t}}\right)e^{-\frac{1}{2}(\gamma-\xi)^2} {\rm sign}\gamma d\gamma}{\sqrt{2\pi}(1\pm\xi/\sqrt{t})}
\end{equation}

Now, we can compute the bid-ask spread
\begin{eqnarray}
    a_{t+1}-b_{t+1} &\simeq& \theta \sqrt{\frac{2}{\pi t}}\frac{1}{1-\frac{\xi^2}{t}}\int_{-\infty}^\infty (\gamma-\xi) e^{-\frac{1}{2}(\gamma-\xi)^2} {\rm sign}\gamma d\gamma \nonumber \\
    &=& 2\theta\sqrt{\frac{2}{\pi t}}e^{-\frac{\xi^2}{2}}+O(t^{-\frac{3}{2}})
\end{eqnarray}

Its expected value is
\begin{equation}
    \E{a_{t+1}-b_{t+1}} = \frac{2\theta}{\sqrt{\pi t}}e^{-\frac{\nu^2t}{4}}=\frac{2\theta}{\sqrt{\pi t}}+O(\nu^2t)
\end{equation}

\end{document}